\documentclass[useAMS,onecolumn]{mn2e}
\usepackage{psfig}
\newif\ifAMStwofonts

\title[The optimal filters for the construction of the ensemble pulsar time]
{The optimal filters for the construction of the ensemble pulsar time}
\author[Alexander E. Rodin]
       {Alexander E. Rodin $^1$\\
      $^1$ Pushchino Radio Astronomical Observatory, Pushchino, Moscow region,
   142290, Russia}
\date{Accepted.............200... ;
      Received ............200... ; 
      in original form ...........200... }
\pubyear{2007}

\def\LaTeX{L\kern-.36em\raise.3ex\hbox{a}\kern-.15em
    T\kern-.1667em\lower.7ex\hbox{E}\kern-.125emX}
\begin{document}
\Large
\label{firstpage}
\baselineskip=20pt

\maketitle

\begin{abstract}
The algorithm of the ensemble pulsar time based on the optimal Wiener
filtration method has been constructed. This algorithm allows the
separation of the contributions to the post-fit pulsar timing
residuals of the atomic clock and pulsar itself. Filters were designed
with the use of the cross- and autocovariance functions of the timing
residuals. The method has been applied to the timing data of
millisecond pulsars PSR B1855+09 and PSR B1937+21 and allowed the
filtering out of the atomic scale component from the pulsar
data. Direct comparison of the terrestrial time TT(BIPM06) and the
ensemble pulsar time PT$_{\rm ens}$ revealed that fractional
instability of TT(BIPM06)--PT$_{\rm ens}$ is equal to
$\sigma_z=(0.8\pm 1.9)\cdot 10^{-15}$. Based on the $\sigma_z$
statistics of TT(BIPM06)--PT$_{\rm ens}$ a new limit of the energy
density of the gravitational wave background was calculated to be
equal to $\Omega_g h^2 \sim 3\cdot 10^{-9}$.
\end{abstract}

\section{Introduction}

The discovery of pulsars in 1967 \cite{Hewish68} showed clearly that
their rotational stability allowed them to be used as astronomical
clocks. This became even more obvious after discovery of the
millisecond pulsar PSR B1937+21 in 1982 \cite{Backer82}. Now a typical
accuracy of measuring the time of arrivals (TOA) of millisecond pulsar
pulses comprises a few microseconds and even hundreds of nanoseconds
for some pulsars. For the observation interval in the order $10^8$
seconds this accuracy produces a fractional instability of $10^{-15}$,
which is comparable to the fractional instability of atomic
clocks. Such a high stability cannot but used for time metrology and
time keeping.

There are several papers considering applicability of stability of
pulsar rotation to time scales. The paper \cite{Guinot88} presents
principles of the establishment of TT (Terrestrial Time) with the main
conclusion that one cannot rely on the single atomic standard before
authorised confirmation and, for pulsar timing, one should use the
most accurate realisations of terrestrial time TT(BIPMXX) (Bureau
International des Poids et Mesures). The paper of Ilyasov
et. al. \cite{ilyasov89} describes the principles of a pulsar time
scale, definition of "pulsar second" is presented. Guinot \& Petit
(1991) show that, because of the unknown pulsar period and period
derivative, rotation of millisecond pulsars is only useful for
investigations of time scale stability a posteriori and with long data
spans. The papers \cite{ilyasov96}, \cite{kopeikin97}, \cite{rodin97},
\cite{ilyasov98} suggest a binary pulsar time (BPT) scale based on the
motion of a pulsar in a binary system with theoretical expressions for
variations in rotational and binary parameters depending on the
observational interval. The main conclusion is that BPT at short
intervals is less stable than the conventional pulsar time scale (PT),
but at a longer period of observation ($10^2\div 10^3$ years) the
fractional instability of BPT may be as accurate as $10^{-14}$. The
paper of Petit \& Tavella \cite{petit96} presents an algorithm of a
group pulsar time scale and some ideas about the stability of BPT. The
paper of Foster \& Backer \cite{foster90} presents a polynomial
approach for describing clock \& ephemeris errors and the influence of
gravitational waves passing through the Solar system.

In this paper the author presents a method of obtaining corrections of
the atomic time scale relative to PT from pulsar timing
observations. The basic idea of the method was published earlier in
the paper \cite{rodin06}.

In Sect.~\ref{sect2}, the main formulae of pulsar timing are
described. Sect.~\ref{sect3} contains theoretical algorithm of Wiener
filtering. Sect.~\ref{sect4} presents the results of numerical
simulation, i.e. recovery of harmonic signal from noisy data by Wiener
optimal filter and weighted average algorithm. The latter one is used
similarly to the paper of Petit \& Tavella \cite{petit96}.
Sect.~\ref{sect5} describes an application of the algorithm to real
timing data of pulsars PSR B1855+09 and PSR B1937+21 \cite{kaspi94}.

\section{Pulsar timing}\label{sect2}

The algorithm of the pulsar timing is widely described in the literature
\cite{BackerHellings86}, \cite{oleg90}, \cite{edwards06}. Two
basic equations are presented below. Expression for the pulsar rotational
phase $\phi(t)$ can be written in the  following form:
\begin{equation}\label{phase}
\phi(t)=\phi_0+\nu t +\frac12\dot\nu{t}^2+\varepsilon(t),
\end{equation}
where $t$ is the barycentric time, $\phi_0$ is the initial phase at
epoch $t=0$, $\nu$ and $\dot\nu$ are the pulsar spin frequency and its
derivative respectively at epoch $t=0$, and $\varepsilon(t)$ is the
phase variations (timing noise). Based on the eq.~(\ref{phase}) pulsar
rotational parameters $\nu$ and $\dot\nu$ can be determined.

The relationship between the time of arrival of the same pulse to the
Solar system barycentre $t$ and to observer site $\hat t$ can be
described by the following equation \cite{murray86}
\begin{equation}\label{tmt}
c(\hat t-t)=-({\bf k}\cdot {\bf b})+\frac{1}{2R}[{\bf k}\times {\bf
b}]^2+\Delta t_{\rm rel}+\Delta t_{DM},
\end{equation}
where ${\bf k}$ is the barycentric unit vector directed to the pulsar,
${\bf b}$ is radius-vector of the radio telescope, $R$ is a distance to
the pulsar, $\Delta t_{\rm rel}$ is the gravitational delay caused by
the space-time curvature, $\Delta t_{DM}$ is the plasma delay. The
pulsar coordinates, proper motion and distance are obtained from the
eq.~(\ref{tmt}) by fitting procedure that includes adjustment of
above-mentioned parameters for minimising the weighted sum of squared
differences between $\phi(t)$ and the nearest integer.


\section{Filtering technique}\label{sect3}

Let us consider $n$ uniform measurements of a random value (post-fit
timing residuals) ${\bf r}=(r_1,r_2,\ldots,r_n)$. ${\bf r}$ is a sum
of two uncorrelated values ${\bf r}={\bf s}+\varepsilon$, where ${\bf
s}$ is a random signal to be estimated and associated with clock
contribution, $\varepsilon$ is random error associated with
fluctuations of pulsar rotation. Both values ${\bf s}$ and
$\varepsilon$ should be related to the {\it ideal} time scale since 
pulsars in the sky "do not know" about time scales used for their
timing. The problem of Wiener filtration is concluded in
estimation of the signal ${\bf s}$ if measurements ${\bf r}$ and
covariances (\ref{cov}) are given \cite{wiener49,gubanov97}. For ${\bf
r}$, ${\bf s}$ and $\varepsilon$ their covariance functions can be
written as follows
\begin{equation}\label{cov}
\begin{array}{l}
{\rm cov}(r,r)=\left<r_i,r_j\right>= \left<s_i,s_j\right>+
\left<\varepsilon_i, \varepsilon_j\right>,\\ 
{\rm cov}(s,s)=\left<s_i,s_j\right>,\\ 
{\rm cov}(s,r)=\left<s_i,r_j\right>=\left<r_i,s_j\right>
= \left<s_i,s_j\right>,\\ 
{\rm cov}(\varepsilon,\varepsilon)=\left<\varepsilon_i,
\varepsilon_j\right>.
\end{array}(i,j=1,2,\ldots,n)
\end{equation}
$\left< \right>$ denominates the ensemble average.

The optimal Wiener estimation of the signal ${\bf s}$ and a posteriori
estimation of its covariance function ${\bf D}_{ss}$ are expressed by
formulae \cite{wiener49,gubanov97}
\begin{equation}\label{signal}
{\bf\hat s}={\bf Q}_{sr}{\bf Q}_{rr}^{-1}{\bf r}
={\bf Q}_{ss}{\bf Q}_{rr}^{-1}{\bf r}
={\bf Q}_{ss}({\bf Q}_{ss}+{\bf Q}_{\varepsilon \varepsilon})^{-1}{\bf
r}
\end{equation}
and
\begin{equation}\label{dss}
{\bf D}_{ss}={\bf Q}_{ss}-{\bf Q}_{sr}{\bf Q}_{rr}^{-1}{\bf Q}_{rs},
\end{equation}
where covariance matrices ${\bf Q}_{rr}$, ${\bf Q}_{sr}$, ${\bf
Q}_{rs}$, ${\bf Q}_{ss}$ are constructed as Toeplitz matrices from the
corresponding covariances.  In this paper we assume that processes
${\bf s}$ and $\varepsilon$ are stationary in a weak sense (stationary
are the first and second moments). Since quadratic fit eliminates the
non-stationary part of a random process \cite{kopeikin99}, their
covariance functions depend on the difference of the time moments
$t_i-t_j$.

Since it is impossible to perform pulsar timing observations without a
reference clock, for separation of the covariances,
$\left<s_i,s_j\right>$ and $\left<\varepsilon_i,
\varepsilon_j\right>$, it is necessary to observe at least two pulsars
relative to the same time scale. In this case, combining the pulsar
TOA residuals and accepting that cross-covariances
$\left<{}^2\varepsilon_i,{}^1\varepsilon_j\right> =
\left<{}^1\varepsilon_i,{}^2\varepsilon_j\right>=0$ produces
(hereafter upper-left indices run over pulsars under consideration)
\begin{equation}
\left<s_i,s_j\right> =
\left(\left<{}^1r_i+{}^2r_i,{}^1r_j+{}^2r_j\right> -
\left<{}^1r_i-{}^2r_i,{}^1r_j-{}^2r_j\right>\right)/4,
\;{\rm or}\;
\left<s_i,s_j\right> = \left<{}^1r_i,{}^2r_j\right>.
\end{equation}
If $M$ pulsars are used for construction of the pulsar time scale then
one has $M(M-1)/2$ signal covariance estimates $\left<s_i,s_j\right> =
\left<{}^k r_i,{}^l r_j\right>,\; (k,l=1,2,\ldots,M)$.

In formula (\ref{signal}) the matrix ${\bf Q}_{rr}^{-1}$ serves as the
whitening filter. The matrix ${\bf Q}_{ss}$ forms the signal from the
whitened data.

The ensemble signal (pulsar time scale) is expressed as follows
\begin{equation}\label{ensemble}
{\bf\hat s_{\rm ens}} = \frac{2}{M(M-1)}
\sum_{m=1}^{{M(M-1)\over 2}}{}_m{\bf Q}_{ss}\cdot
\sum_{i=1}^{M}{}{^i}w\;{}^i{\bf Q}_{rr}^{-1} \cdot{}^i{\bf r},
\end{equation}
where $^i w$ is the relative weight of the $i$th pulsar, $^iw=
\kappa/\sigma_i^2$, $\sigma_i$ is the root-mean-square of whitened
data ${}^i{\bf Q}_{rr}^{-1} \cdot{}^i{\bf r}$, and $\kappa$ is the
constant serving to satisfy $\sum_i\, {^iw}=1$. The first multiplier
in eq.(\ref{ensemble}) is the average cross-covariance function, the
second multiplier is the weighted sum of the whitened data.

For calculation of the auto- and cross-covariances, the following
algorithm was used: the initial time series $^kr_t$ were
Fourier-transformed
\begin{equation}
{}^kx(\omega)=\frac{1}{\sqrt{n}}\sum_{t=1}^n {}^kr_t h_t
e^{i\omega t},\;(k=1,2,\ldots,M),
\end{equation}
weights $h_t$ are the 0th order discrete prolate spheroidal sequences
 \cite{percival91} which are used for optimisation of broad-band bias
 control. These can be calculated to a very good approximation using
 the following formula \cite{percival91}
\begin{equation}
h_t=C_0'\frac{I_0\left(\pi W(n-1)\sqrt{\left[1-(\frac{2t-1}{n}-1)^2
\right]} \right)}{I_0(\pi W(n-1))},
\end{equation}
where $C_0'$ is the scaling constant used to force the convention
$\sum h_t^2=1$, $I_0$ is the modified Bessel function of the 1st kind
and 0th order, and the parameter $W$ affects the magnitude of the
side-lobes in the spectral estimates (usually $W=1 \div 4$). In this
paper $W=1$ was used.

Power spectrum ($k=l$) and cross-spectrum ($k\neq l$) were calculated
by the formula
\begin{equation}\label{spxy}
{}^{kl}X(\omega)=\frac{1}{2\pi}\left|^kx(\omega)^lx^{*}(\omega)
\right|,\end{equation}
where $(\cdot)^*$ denominates complex conjugation.

Lastly, auto- ($k=l$) and cross-covariance ($k\neq l$) were calculated
using the following formula
\begin{equation}
{\rm cov}(^kr,^lr) =\sum_{\omega=1}^n {^{kl}}X(\omega)
e^{-i\omega t},\;(k,l=1,2,\ldots,M).
\end{equation}

\section{Computer simulation}

\begin{figure}
\centering{
\hspace*{-1.cm}
\vbox{\hbox{\hspace*{4.4cm}(a)}\psfig{figure=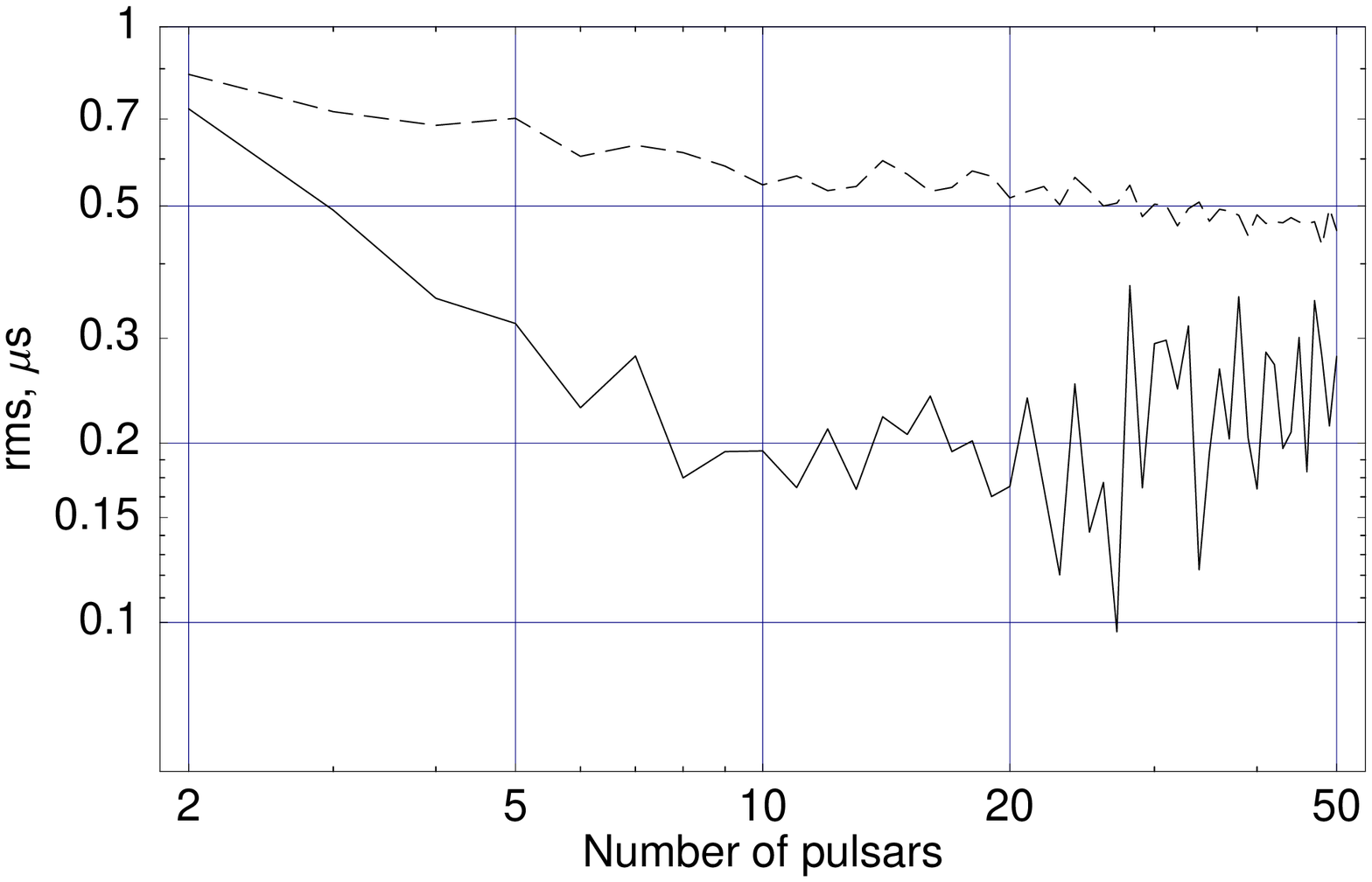,width=8cm}} 
\vbox{\hbox{\hspace*{4.4cm}(b)}\psfig{figure=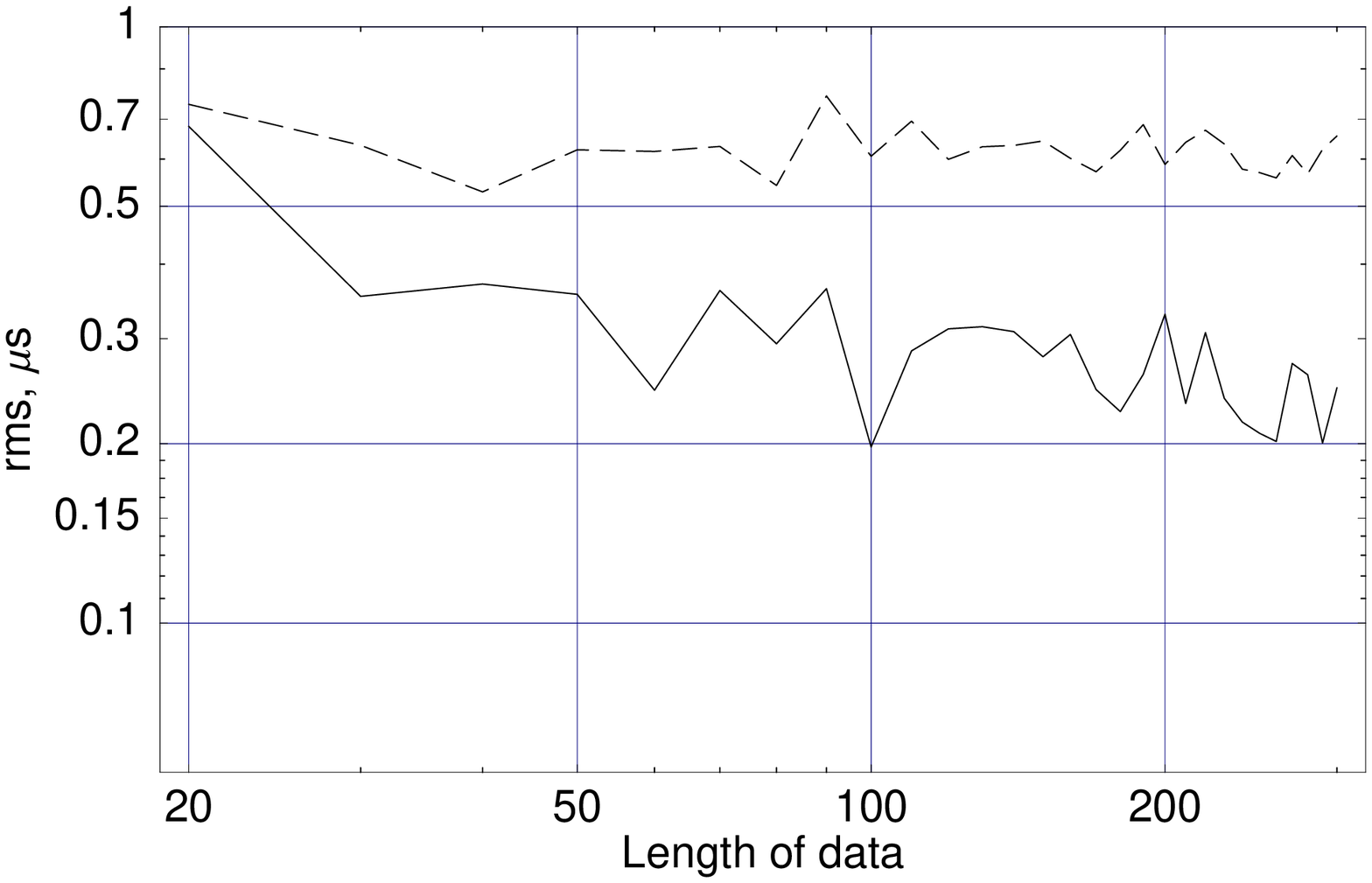,width=8cm}} 
}
\vspace{3mm}
\centering{
\hspace*{-1.cm}
\vbox{\hbox{\hspace*{4.4cm}(c)}\psfig{figure=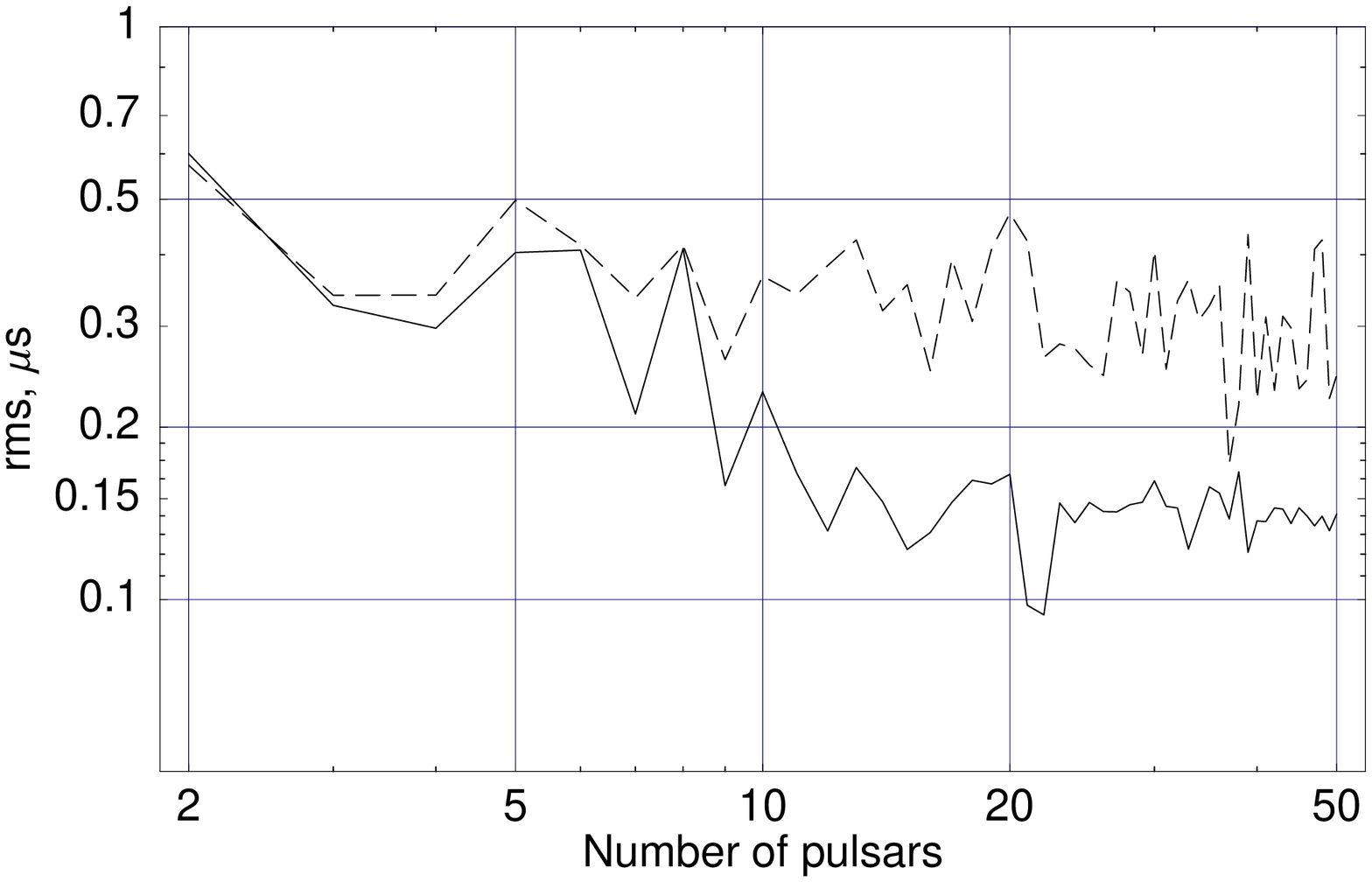,width=8cm}}
\vbox{\hbox{\hspace*{4.4cm}(d)}\psfig{figure=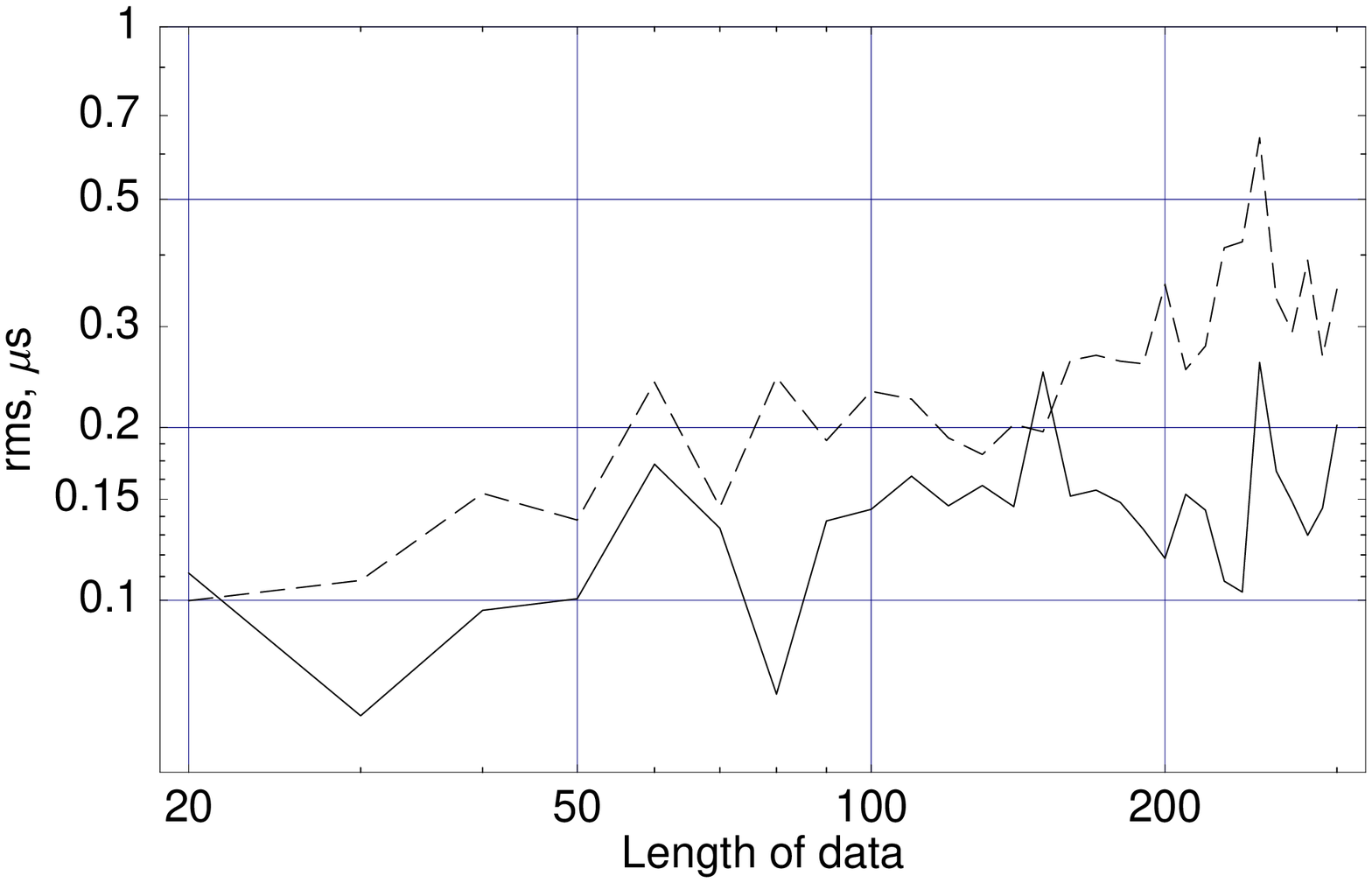,width=8cm}}
}
\vspace{3mm}
\centering{
\hspace*{-1.cm}
\vbox{\hbox{\hspace*{4.4cm}(e)}\psfig{figure=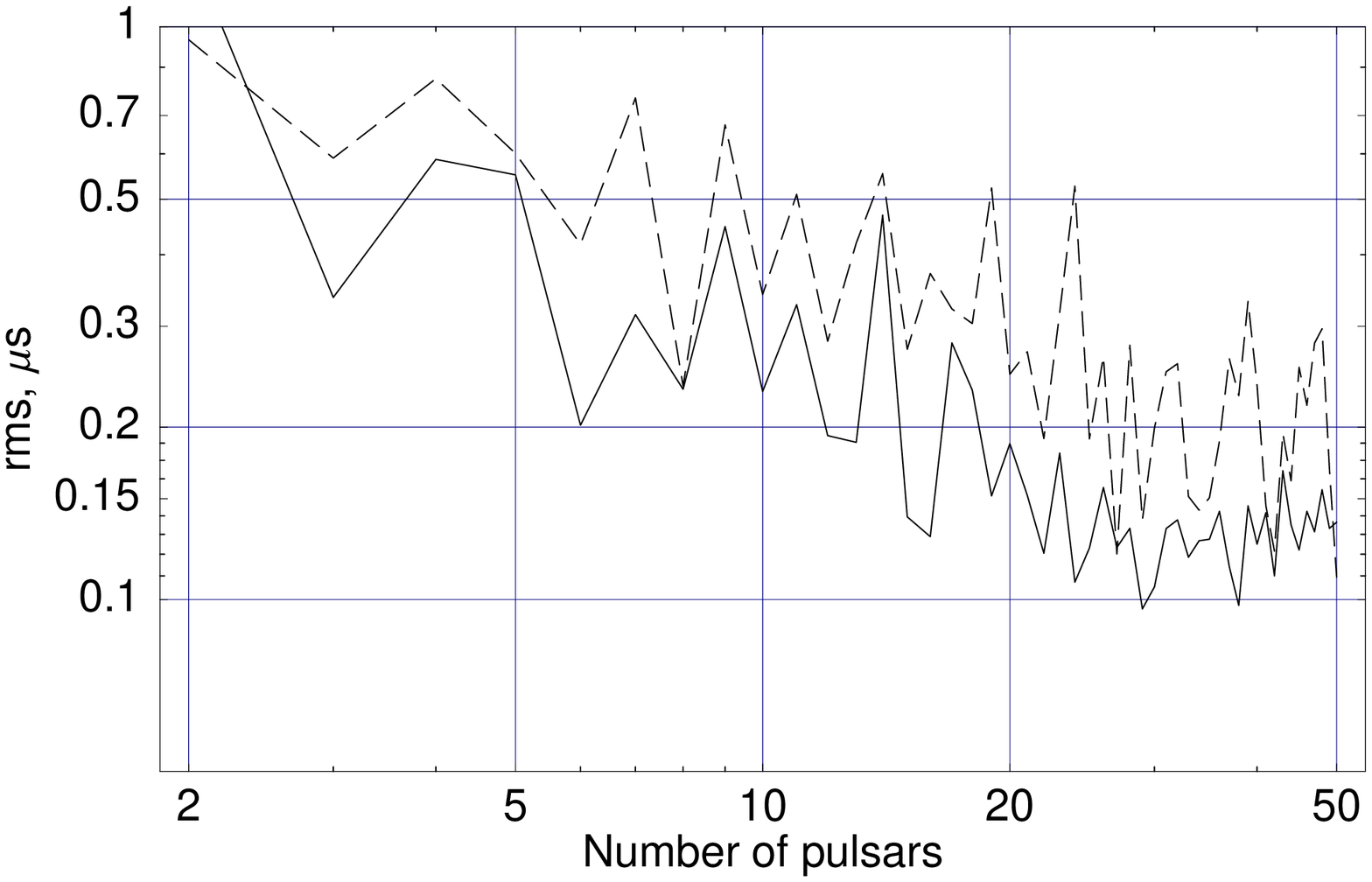,width=8cm}}
\vbox{\hbox{\hspace*{4.4cm}(f)}\psfig{figure=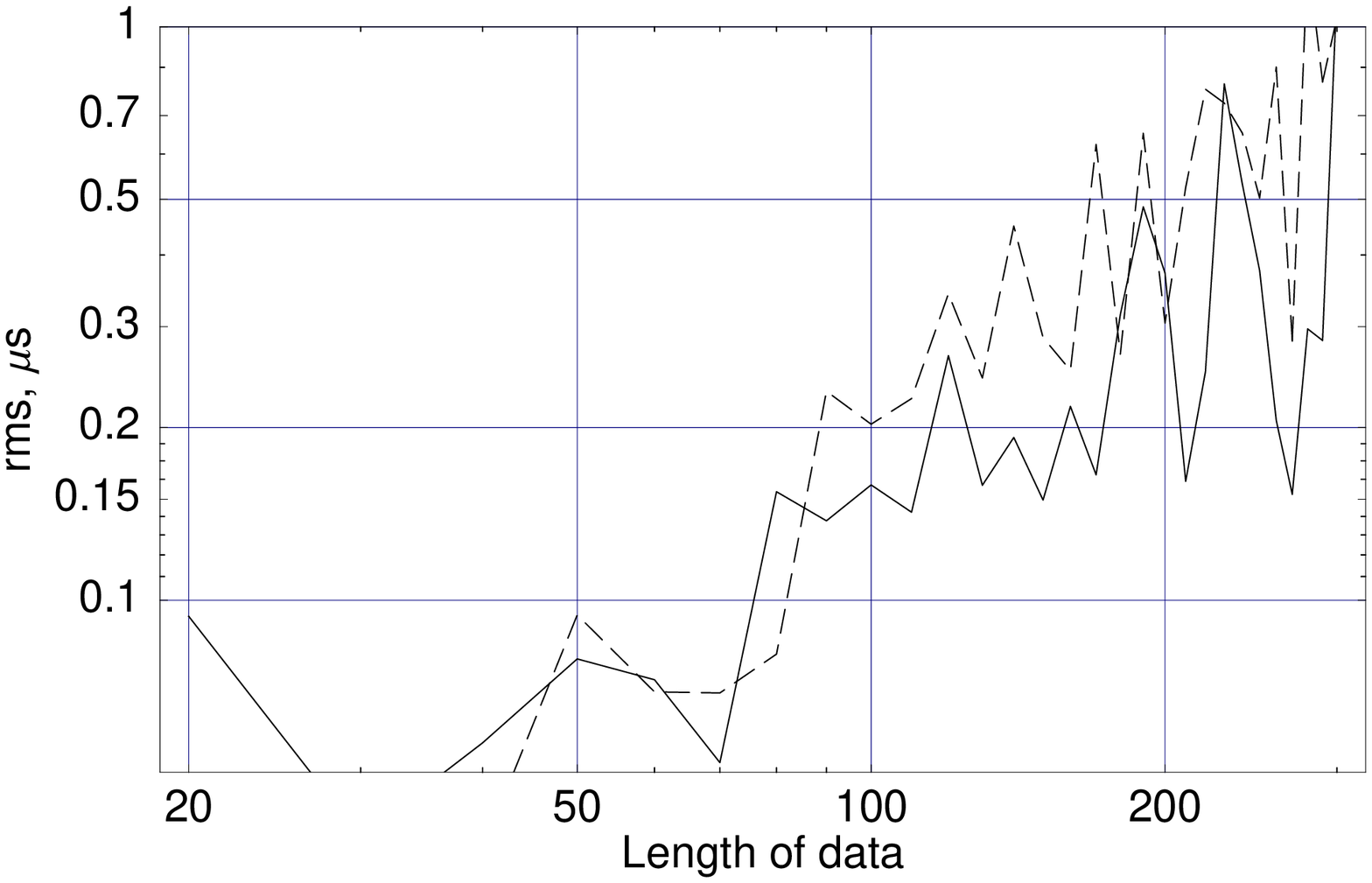,width=8cm}}
}
\caption{The accuracy of signal estimation based on the methods of
weighted average (dashed line) and Wiener filter (solid line) as
dependent on the number of pulsars (left panels) and length of the
data (right panels). For the calculation shown in the left panels 256
points of data were taken, for calculations shown in the right panels
five pulsars were used. Different types of noise were generated: (a),
(b) - white phase noise, (c), (d) - white noise in frequency, (e), (f)
- random walk noise in frequency. Data in the panels (d) and (f) were
scaled accordingly for fitting within in all the panels.}
\label{fig1}
\end{figure}

To evaluate performance of the Wiener filtering method as compared to
the weighted average method, we have applied it to simulated time
sequences corresponding to harmonic signal with additive white and red
(correlated) noise. Simulated time series were generated with the help
of random generator built in the {\it Mathematica} software which had
the preset (normal) distribution for different numbers of pulsars. A
maximum of 50 pulsars were used (limited by acceptable computing
time). The harmonic signal to be estimated was as follows:
$A\sin(t/P)$, $A=1$, $P=10$, $t=1,2,\ldots, 256$. The additive
Gaussian white noise had zero mean and different variance. For
example, in simulation for 50 pulsars, the variance was in the range
of $\sigma^2=1,2,\ldots,50$.  The correlated noise $n_2, n_4$ with the
power spectra $1/f^2$ and $1/f^4$ was generated as a single or twice
repeated cumulative sum of the white noise $n_0$:
\begin{equation}
n_{2j}=\sum_{i=1}^{j}n_{0i},\; n_{4j}=\sum_{i=1}^{j}n_{2i},\;
(j=1,2,\ldots,n).
\end{equation}
The second order polynomial trend was subtracted from the generated
time series $n_2$ and $n_4$. The weights of the individual time series
were taken inversely proportional to $\sigma_z^2$, where $\sigma_z$ is
the fractional instability \cite{taylor91}. Quality of the two methods
was compared by calculating the root mean square of the difference
between original and recovered signals.

Fig.~\ref{fig1} shows the results of computer simulation described
above. Quality of these two methods of signal estimation is clearly
visible. It is important to note that signal estimation accuracy in
the case of Wiener filter (solid line) is better in all cases. The
most significant advantage of the Wiener filter over the weighted
average method is seen in the case of the white noise (fig.~1(a),
1(b)). For the correlated noise with the power spectrum $1/f^2$
(fig.~1(c), 1(d)) the advantage is still clear. For the red noise with
the power spectrum $1/f^4$ both methods show similar results
(fig.~1(e), 1(f)). Noteworthy is dependence of estimation quality on
the observation interval for the correlated noise (fig.~1(d), 1(f)):
as the observation interval increases, the estimation accuracy
grows. Physically such a behaviour corresponds to appearance of more
and more strong variations of the correlated noise with time, which
deteriorate the signal estimation quality. Influence of the form of
the correlated noise and length of the observation interval on the
variances of the pulsar timing parameters are described in detail in
\cite{kopeikin97}.

\section{Results}\label{sect4}

To evaluate the performance of the proposed Wiener filter method,
timing data of pulsars PSR B1855+09 and B1937+21 \cite{kaspi94} were
used. For the sake of simplicity of the subsequent matrix
computations, unevenly spaced data were transformed into uniform ones
with a sampling interval of 10 days by means of linear
interpolation. Such a procedure perturbs a high-frequency component of
the data while leaving a low-frequency component, of most interest to
us, unchanged. The sampling interval of 10 days was chosen to preserve
approximately the same volume of data.

\begin{figure*}
\centering{
\hspace*{-1.cm}
\vbox{\hbox{\hspace*{4.4cm}(a)}\psfig{figure=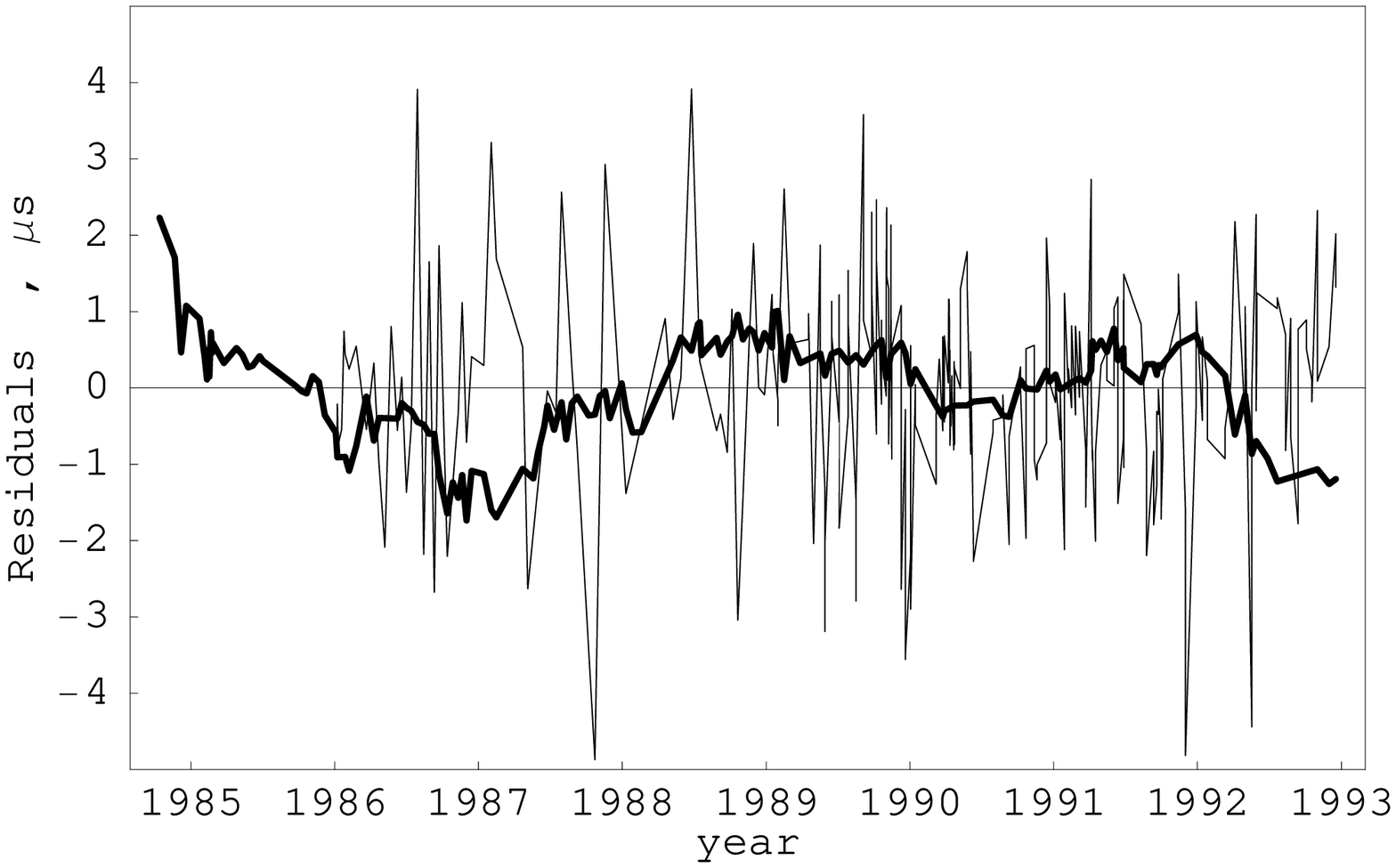,width=8cm}} 
\vbox{\hbox{\hspace*{4.4cm}(b)}\psfig{figure=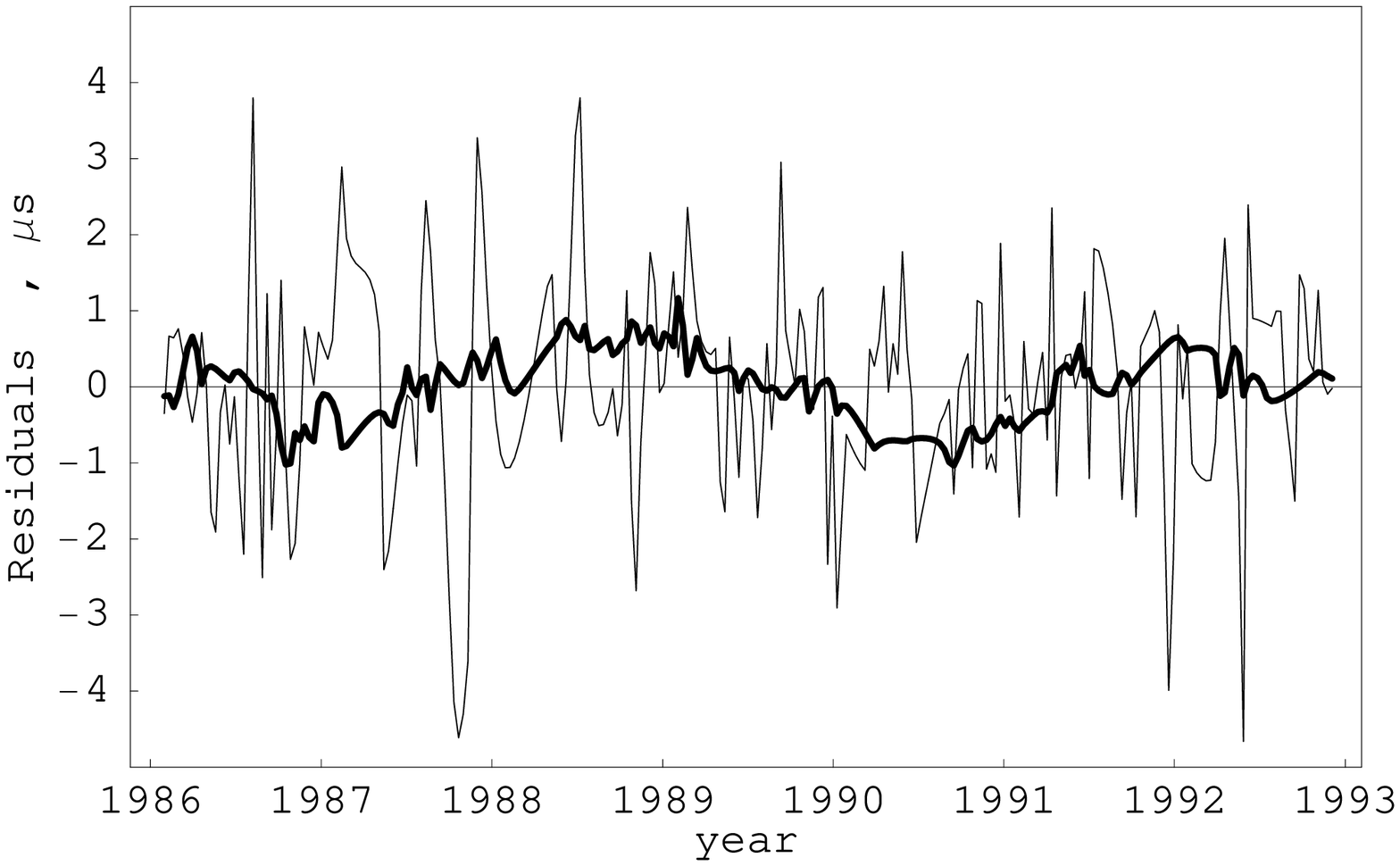,width=8cm}} 
}
\caption{The barycentric timing residuals of pulsars PSR B1855+09
(thin line) and PSR B1937+21 (solid line) before (a) and after (b)
uniform sampling.}
\label{res12}
\end{figure*}

The common part of the residuals for both pulsars (251 TOAs) has been
taken within the interval $MJD=46450\div 48950$. Since choosing the
common part of the time series changes the mean and slope, the
residuals have been quadratically refitted for consistency with the
classical timing fit. The pulsar post-fit timing residuals before and
after processing described above are shown in Fig.~\ref{res12}.

According to \cite{kaspi94}, the timing data of PSRs B1855+09 and
B1937+21 are in UTC time scale. UTC (Universal Coordinated Time) is
the international atomic time scale that serves as the basis for
timekeeping for most of the world. UTC runs at the same frequency as
TAI (International Atomic Time). It differs from TAI by an integral
number of seconds. This difference increases when so-called leap
seconds occur. The purpose of adding leap seconds is to keep atomic
time (UTC) within $\pm 0.9$~s of an time scale called UT1, which is
based on the rotational rate of the Earth. Local realisations of UTC
exist at the national time laboratories. These laboratories
participate in the calculation of the international time scales by
sending their clock data to the BIPM. The difference between UTC
(computed by BIPM) and any other timing centre's UTC only becomes
known after computation and dissemination of UTC, which occurs about
two weeks later. This difference is presently limited mainly by the
long-term frequency instability of UTC \cite{audoin01}.

\begin{figure*}
\centering{
\hspace*{-1.cm}
\vbox{\hbox{\hspace*{4.4cm}(a)}\psfig{figure=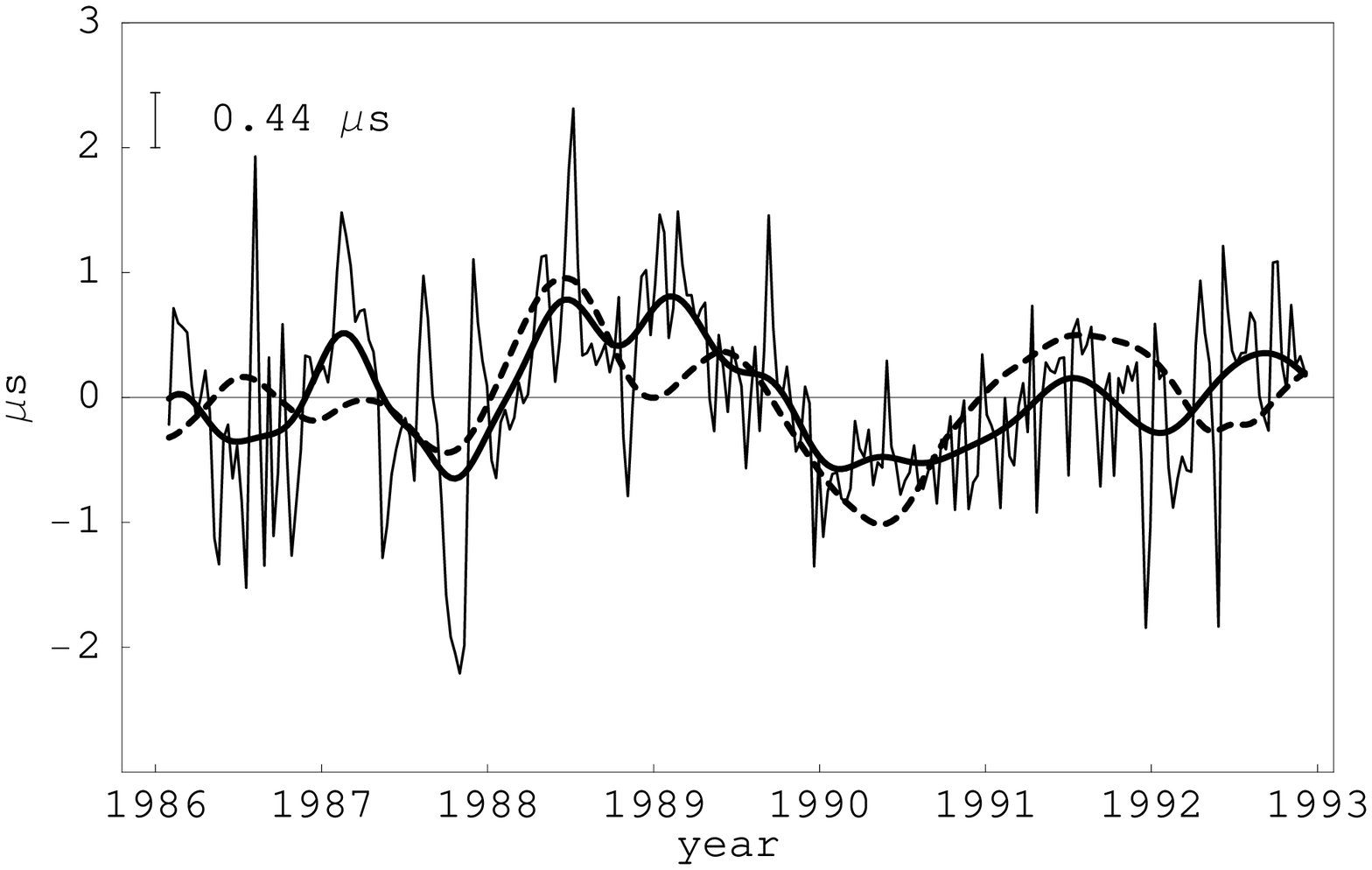,width=8cm}} 
\vbox{\hbox{\hspace*{4.4cm}(b)}\psfig{figure=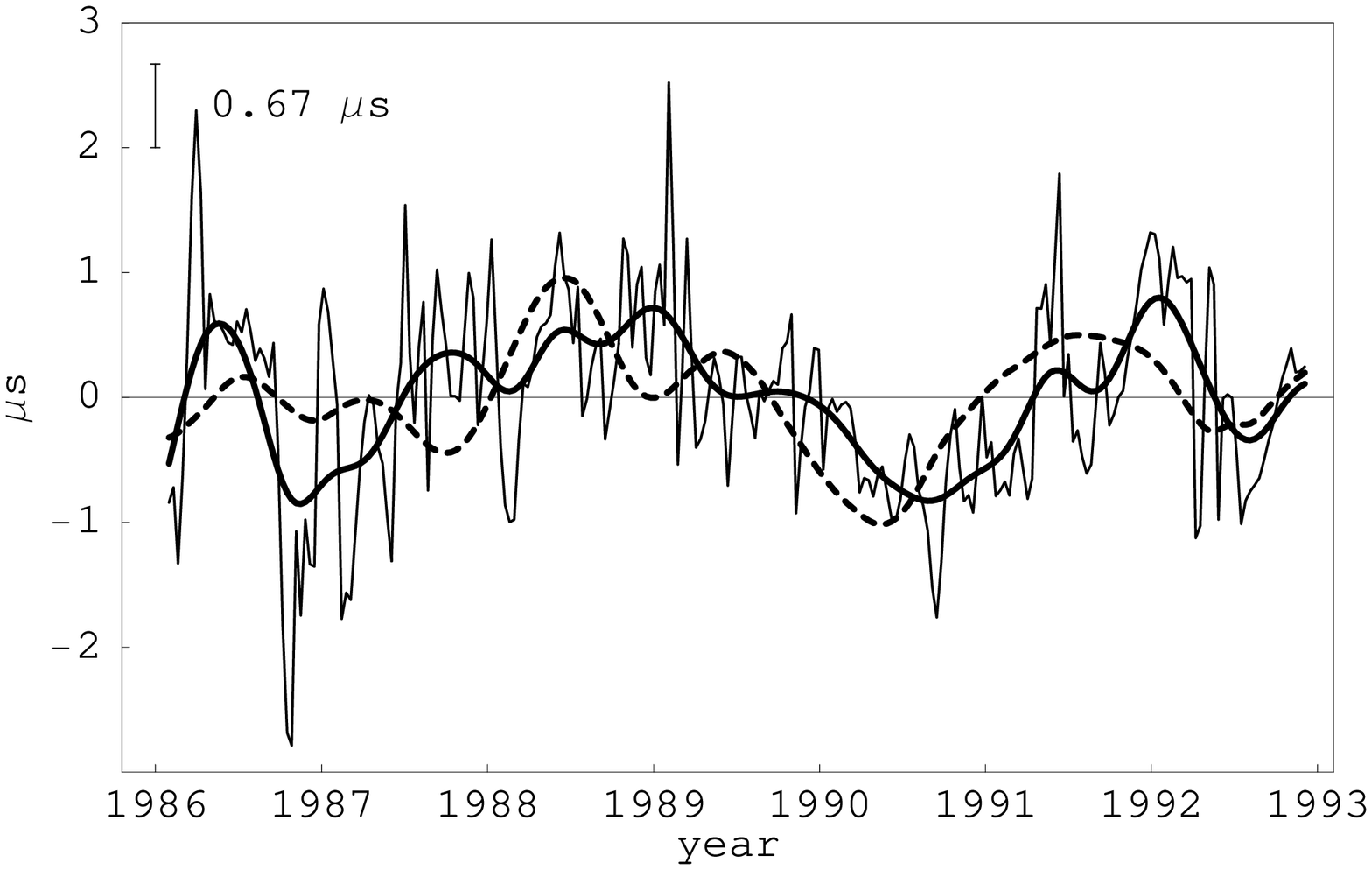,width=8cm}}} 
\caption{ Differences in UTC -- PT$_{1855}$ (a) and UTC -- PT$_{1937}$
(b) (thin line) for the interval $MJD=46450\div48950$ estimated using
the optimal filtering method (eq.~(4)). The smoothing with Kaiser
filter is shown by solid line. The dashed line displays the difference
UTC -- TT(BIPM06).}
\label{signal12}
\end{figure*}

\begin{figure*}
\centering \psfig{width=10cm,figure=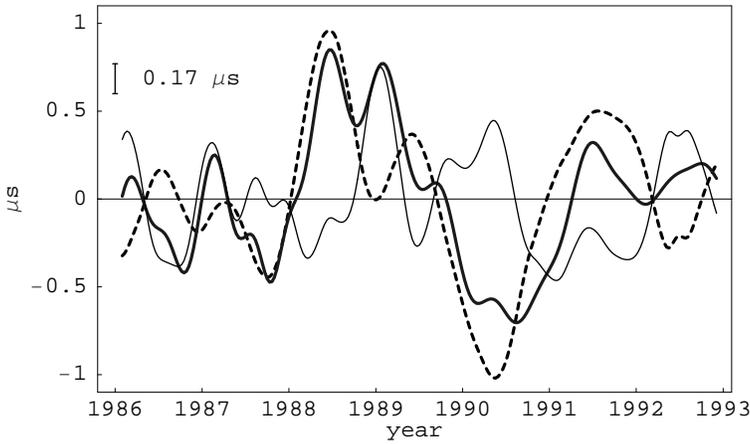}
\caption{Combined clock variations of UTC -- PT$_{\rm ens}$ for the
interval $MJD=46450\div48950$ estimated using the optimal filtering
method from the timing residuals of pulsars PSR B1855+09 and PSR
B1937+21 (solid line) and difference UTC -- TT(BIPM06) (dashed line).
The thin line indicates the difference TT -- PT$_{\rm ens}$}
\label{coll-est}
\end{figure*}

The signal we need to estimate is the difference UTC --
PT. Fig.~\ref{signal12} shows the signal estimates (thin line) based
on the timing residuals of pulsars PSR B1855+09 and PSR B1937+21 and
calculated with use of eq.~(\ref{signal}). The combined signal
(ensemble pulsar time scale, eq.~(\ref{ensemble})) is shown in
fig.~\ref{coll-est} and displays behaviour similar to the difference
UTC -- TT(BIPM06) (correlation $\rho=0.75$).

All three signals UTC -- PT$_{1855}$, UTC -- PT$_{1937}$ and UTC --
PT$_{\rm ens}$ were smoothed by use of the following method: to
decrease the Gibbs phenomenon (signal oscillations) near the ends of
smoothing interval, the series under consideration were backward and
forward forecasted by $p=IntegerPart[n/2]$ lags ($n=251$ is length of
time series) with use of the Burg's (also referred to as the maximum
entropy method) autoregression algorithm of order $p$
\cite{burg75,terebizh92}. A new time series of double length were
smoothed by use of the low-pass Kaiser filter \cite{gold73,kaiser74}
with the bandwidth of $f_{\rm max}/32$, where $f_{\rm max}=2/\Delta
t$, $\Delta t = 10$ days is the sampling interval. The choice of the
bandwidth was defined by visual comparison with the UTC -- TT(BIPM06)
line. The final time series were obtained by dropping the forward and
backward forecasted sections of the smoothed series.

The accuracy of the obtained realisations of the pulsar time UTC --
PT$_{1855}$ and UTC -- PT$_{1937}$ was derived from the diagonal
elements of the covariance matrix defined by eq.~(\ref{dss}).  The
root mean square value of UTC -- PT$_{1855}$ and UTC -- PT$_{1937}$ is
equal to 0.44 $\mu$s and 0.67 $\mu$s respectively. The accuracy of the
smoothed signals was estimated as $0.44/\sqrt{16}=0.11$ $\mu$s and
$0.67/\sqrt{16}=0.17$ $\mu$s. Finally, for overall accuracy a
conservative estimate 0.17 $\mu$s was accepted.

\section{Discussion}\label{sect5}

The stability of a time scale is characterised by so-called Allan
variance numerically expressed as a second-order difference of the
clock phase variations. Since timing analysis usually includes
determination of the pulsar spin parameters up to at least the first
derivative of the rotational frequency, it is equivalent to excluding
the second order derivative from pulsar TOA residuals and therefore
there is no sense in the Allan variance. For this reason, for
calculation of the fractional instability of a pulsar as a clock,
another statistic $\sigma_z$ has been proposed \cite{taylor91}. A
detailed numerical algorithm for calculation of $\sigma_z$ has been
described in the paper \cite{matsakis97}.

In this work, an idea has been proven that different realisations of
pulsar time scales must have comparable stability between each other
\cite {lyne98} and should be not worse than available terrestrial time
scale at the same interval. For this purpose statistic $\sigma_z$ of
two realisations of PT, UTC -- PT$_{1855}$ and UTC -- PT$_{1937}$,
were compared.

\begin{figure*}
\centering \psfig{width=10cm,figure=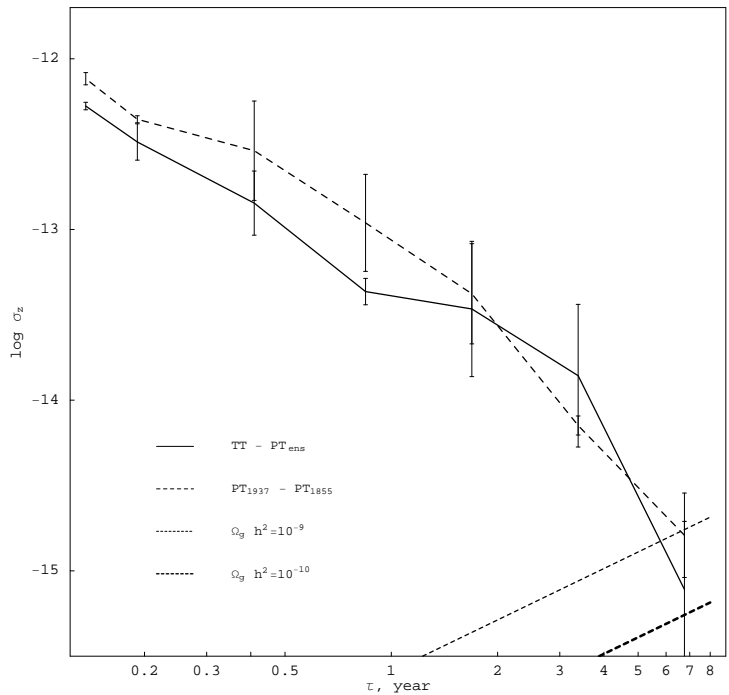}
\caption{The fractional instability $\sigma_z$ based on the difference
PT$_{1937}$--PT$_{1855}$ (dashed line) and $\sigma_z$ of the
difference TT -- PT$_{\rm ens}$ (solid line). Theoretical $\sigma_z$
in the case $\Omega_g h^2=10^{-9}$ and $10^{-10}$ are shown in the
lower right-hand corner of the plot.}
\label{sigma-z}
\end{figure*}

Fig.~\ref{sigma-z} presents the fractional instability of the
differences PT$_{1937}$ -- PT$_{1855}$ (dashed line) and TT --
PT$_{\rm ens}$ (solid line). At 7 years time interval
$\sigma_z=(0.8\pm 1.9)\cdot 10^{-15}$ and $\sigma_z=(1.6\pm 2.9)\cdot
10^{-15}$ for TT -- PT$_{\rm ens}$ and PT$_{1937}$~--~PT$_{1855}$
respectively. One can see that instability of the two differences lays
within error bar intervals. The fractional instability of TT relative
to PT$_{\rm ens}$ and PT$_{1937}$ relative to PT$_{1855}$ is almost
one order of magnitude better than individual fractional instability
of the pulsars PSR B1855+09 and PSR B1937+21.

As an example of astrophysical application of the fractional
instability values obtained in this work, one could consider
estimation of the upper limit of the energy density of the stochastic
gravitational wave background \cite{kaspi94}. For this purpose
theoretical lines of $\sigma_z$ in the case when the gravitational
wave background with the fractional energy density $\Omega_g
h^2=10^{-9}$ and $10^{-10}$ begins to dominate are plotted in the
lower right hand side corner of fig.~\ref{sigma-z}. One can see that
$\sigma_z$ of TT -- PT$_{\rm ens}$ crosses the line $\Omega_g
h^2=10^{-9}$ and approaches $\Omega_g h^2=10^{-10}$. The upper limit
of $\Omega_g h^2$ based on the two sigma uncertainty (95\% confidence
level) of the ensemble $\sigma_z$ is equal to $\sim 3\cdot 10^{-9}$.

Noteworthy, that since PSRs 1855+09 and 1937+21 are relatively close
to each other in the sky (angular separation is $15.5^{\circ}$) and
hence their variations of the rotational phase contain a correlated
contribution caused by the stochastic gravitational wave background
\cite{hellings83}, they form a good pair for estimation of the upper
limit of $\Omega_g h^2$.

Currently, the accuracy of the filtering method without contribution
of the uncertainty of the TT algorithm is estimated at a level of 0.17
$\mu$s.  So, the uncertainty in PT$_{\rm ens}$ may, in principle,
reach the level of a few ten nanoseconds if it were to be used for all
high-stable millisecond pulsars. As computer simulations show, for the
highest advantage while applying the Wiener optimal filters one should
use the pulsars that show no correlated noise in their post-fit timing
residuals.

\section{Conclusions}

An algorithm of forming of ensemble pulsar time scale based on the
method of the optimal Wiener filtering is presented.  The basic idea
of the algorithm consists in the use of optimal filter for removal of
additive noise from the timing data {\it before} construction of the
weighted average ensemble time scale.

Such a filtering approach offers an advantage over the weighted
average algorithm since it utilises additional statistical information
about common signal in the form of its covariance function or power
spectrum. Since timing data are always available relative to a
definite time scale, for separation of the pulsar and clock
contributions one need to use observations from a few pulsars (minimum
two) relative to the same time scale. Such approach allows estimation
of the signal covariance function (power spectrum) by averaging all
cross-covariance functions or power cross-spectra of the original
data.

The availability of two scale differences UTC -- TT and UTC -- PT has
resulted in the long awaited possibility of comparing the ultimate
terrestrial time scale TT and extraterrestrial ensemble pulsar time
scale PT of comparable accuracy. The fractional instability of the
terrestrial time scale TT relative to PT and their high correlation
have demonstrated that PT scale can be successfully used for
monitoring the long-term variations of TT.

\section*{Acknowledgements}
This work is supported in part by the Russian Foundation for Basic
Research, grant RFBR-06-02-16888.

\end{document}